# Insight into Potential Well Based Nanoscale FDSOI MOSFET Using Doped Silicon Tubs- A Simulation and Device Physics Based Study: Part II: Scalability to 10 nm Gate Length

Shruti Mehrotra, S. Qureshi

*Department of Electrical Engineering, Indian Institute of Technology Kanpur, India*

**Abstract**

The doped silicon regions (tubs) in PWFDSOI MOSFET cause significant reduction in OFF current by reducing the number of carriers contributing to the OFF current. The emphasis of the simulation and device physics study on PWFDSOI MOSFET presented in this paper is on the scalability of the device to 10 nm gate length and its related information. A high $I_{ON}/I_{OFF}$ ratio of 7.6 x $10^5$ and subthreshold swing of 87 mV/decade were achieved in 10 nm gate length PWFDSOI MOSFET. The study was performed on devices with unstrained silicon channel.

## 1. Introduction

The planar FDSOI MOSFET not only has a planar topology for easy integration but also has the back bias feature which allows better front gate control than its bulk counterpart. The ultrathin body and BOX (UTBB) and ground plane (GP) features of the FDSOI MOSFET help in the reduction of drain induced barrier lowering (DIBL). These advantages make FDSOI MOSFET more attractive to designers than FinFETs. FinFETs due to their 3D topology have design and processing complexities. A novel planar potential well based FDSOI MOSFET (PWFDSOI MOSFET) 20 nm gate length was proposed recently [1]. The potential wells in the source and drain regions of PWFDSOI MOSFET are instrumental in reducing the OFF current by orders of magnitude. This device has been discussed in detail in Part I. In this manuscript, we discuss in detail the scalability of PWFDSOI MOSFET to 10 nm gate length. The physics of the device, DIBL and effects like tunneling are discussed.

This paper is organised as follows: Section 2 discusses the PWFDSOI MOSFET at 10 nm gate length and the simulation methodology followed to realize the conventional FDSOI MOSFET with ground plane (GP) (reference device in this study) and the proposed PWFDSOI MOSFET. Section 3 discusses the interpretation of results and Section 4 draws the conclusion.

## 2. Scalability to 10 nm Gate Length

The 20 nm gate length PWFDSOI MOSFET discussed in Part I was scaled to 10 nm gate length to explore the scalability of this device to deep submicron nodes. The simulation and device physics study was performed using Silvaco TCAD on devices with unstrained silicon channel [2]. The simulation models invoked to capture the physics of the devices are mentioned in Table 1 and the device parameters used in this simulation study are mentioned in Table 2. The gate-to-source and gate-to-drain overlap was 2 nm. This was achieved through efforts to optimize the $I_{ON}/I_{OFF}$ ratio. The doping of $T_S$ and $T_D$ was increased to 2 x $10^{20}$ $cm^{-3}$. This increase in doping of $T_S$ and $T_D$ was to make the potential wells in source and drain more effective because the leakage current increases as we scale down the technology node. Devices with BOX thickness of 10 nm and 15 nm were studied [3, 4, 5]. Figure 1(a) shows the 10 nm PWFDSOI MOSFET with a BOX thickness of 10 nm. The relative potential profile along the cutline Z-Z' shown in Fig. 1(b) shows the potential wells in the source and drain regions for the OFF state of the device ($V_{DS}$ = 50 mV, $V_{GS}$ = $V_{BS}$ = 0 V).

*Corresponding author
*Email address:* mshruti@iitk.ac.in (Shruti Mehrotra)

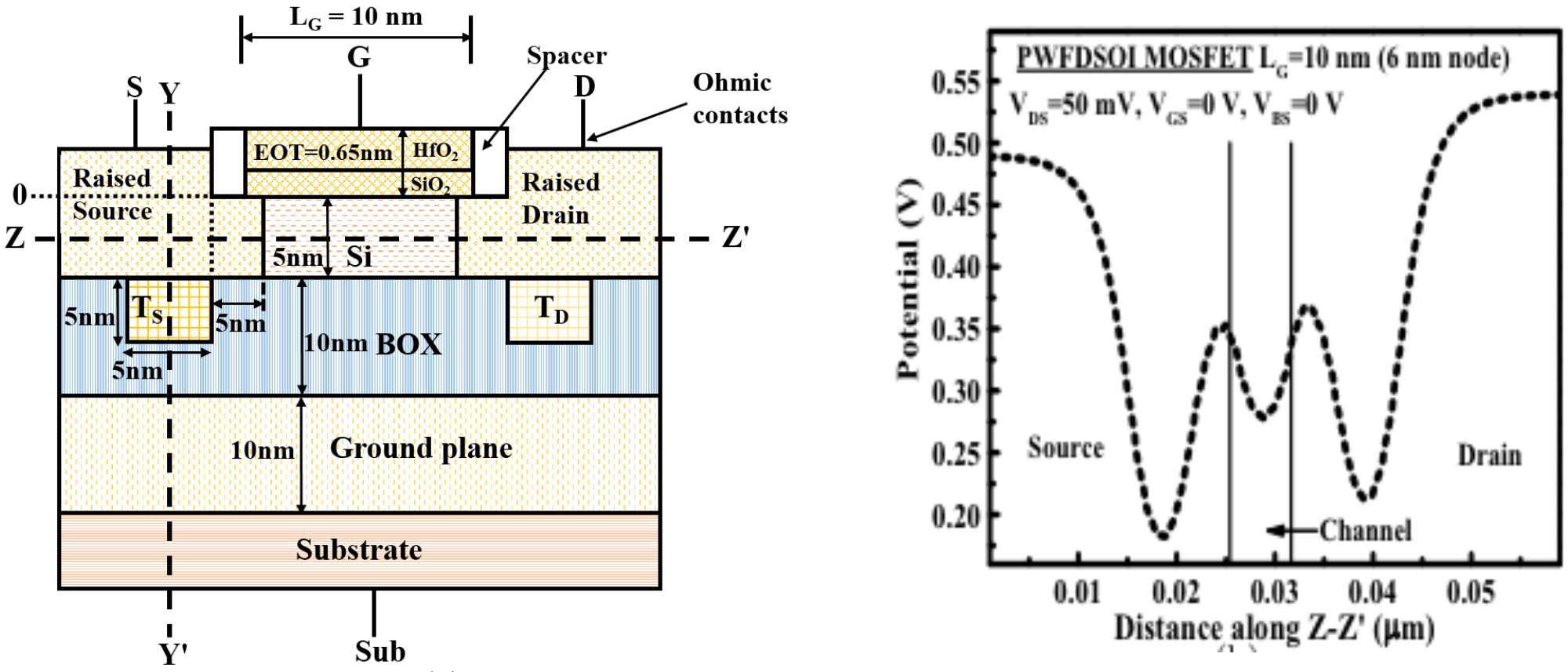

(a) (b)

Fig. 1: (a)10 nm PWFDSOI MOSFET with regions $T_S$ and $T_D$ under the source and drain respectively. The doping of $T_S$ and $T_D$ is p-type with a concentration of 2 x $10^{20}$ $cm^{-3}$. It has an HKMG gate stack with an EOT of 0.65 nm and a BOX thickness of 10 nm. The cutline Z-Z' is through the center of the channel (2.5 nm from the gate stack/channel interface). The vertical cutline Y-Y' is through the center of $T_S$. The term node in this study refers to the effective channel length which in case of 10 nm gate length is 6 nm. (b) Potential wells in source and drain regions in 10 nm PWFDSOI MOSFET with BOX thickness of 10 nm when device is in OFF state ($V_{DS}$ = 50 mV, $V_{GS}$ = $V_{BS}$ = 0 V).

Table 1: Models used in 2D TCAD simulations of FDSOI MOSFET

| TCAD Models | Physical Effect Captured |
|---|---|
| Drift Diffusion | Carrier transport |
| SRH and Auger | Carrier recombination |
| Quantum confinement | SOI layer is very thin leading to quantum confinement of carriers |
| Lombardi mobility model | Acoustic phonon scattering at low fields and surface recombination scattering at high transverse fields |
| High field mobility model | Velocity saturation effect |
| Self heating model | Lattice heating in the SOI layer |
| Fermi Dirac carrier statistics | Presence of heavily doped regions in the device |
| Bandgap narrowing | At very high doping in silicon, the *pn* product becomes doping dependent |

Table 2: Device Parameters used in Simulations at 10 nm Gate Length

| Parameter | Value |
|---|---|
| Gate Length | 10 nm |
| EOT | 0.65 nm |
| $HfO_2$ thickness | 2.2 nm |
| $SiO_2$ thickness | 0.3 nm |
| Permittivity of $HfO_2$ | 25 |
| SOI layer thickness | 5 nm |
| BOX thickness | 10 nm |
| GP thickness | 10 nm |
| Spacer length | 3 nm |
| Gate-to-source/drain overlap | 2 nm |
| SOI layer doping | $10^{15}$ $cm^{-3}$ |
| Source/Drain doping | $10^{20}$ $cm^{-3}$ |
| $T_S$/$T_D$ doping | 2 x $10^{20}$ $cm^{-3}$ |
| Ground plane doping | $10^{20}$ $cm^{-3}$ |
| Substrate doping | $10^{15}$ $cm^{-3}$ |
| Work function of gate metal | 4.52 eV |

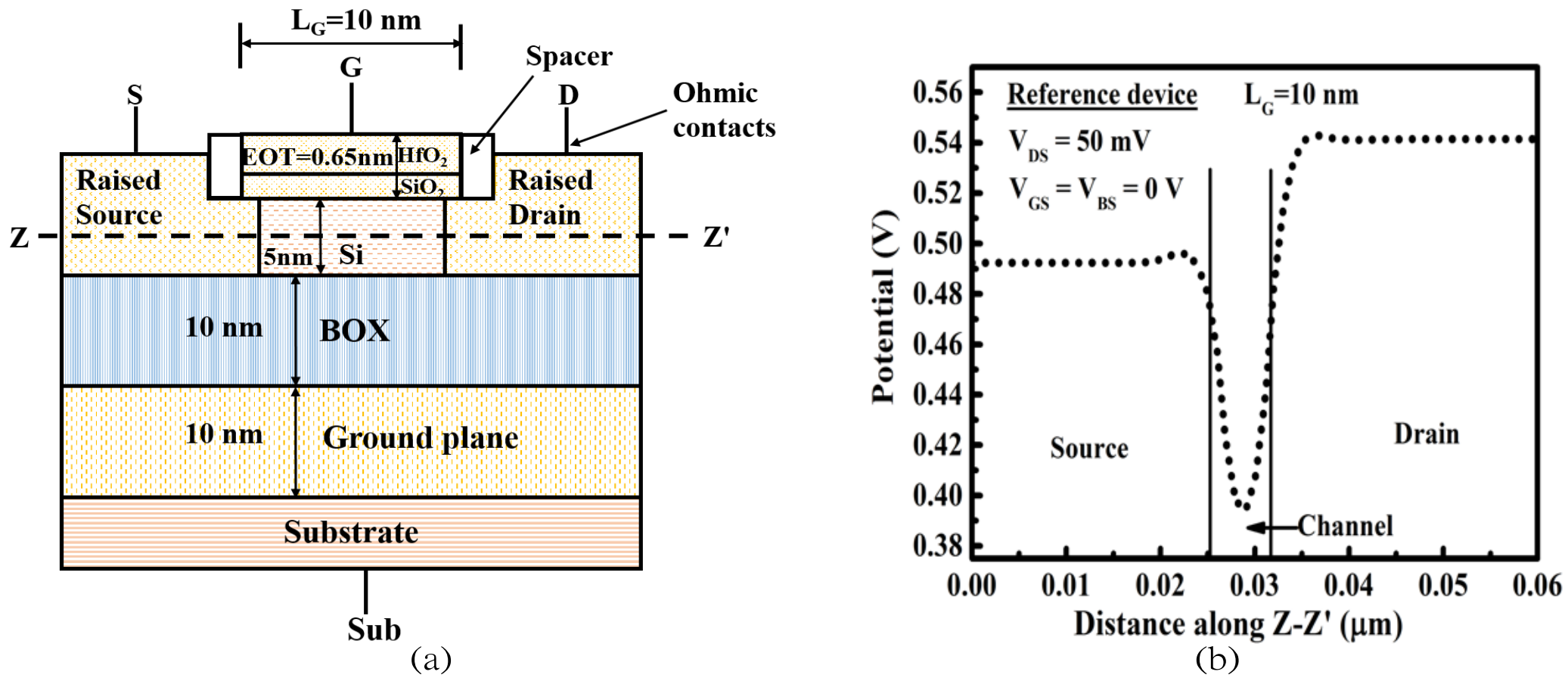

Fig. 2: (a) Schematic of 10 nm FDSOI MOSFET with $p^+$ ground plane (GP) under the BOX. The horizontal cutline Z-Z' is drawn in the center of the channel (2.5 nm from the gate stack/channel interface). (b) Relative channel potential profile in FDSOI MOSFET with GP. Absence of potential wells in source and drain regions in 10 nm FDSOI MOSFET is observed.

The reference device in this study was a 10 nm gate length ground plane (GP) FDSOI MOSFET as shown in Fig. 2. The PWFDSOI MOSFET shown in Fig. 1(a) is identical to the reference device shown in Fig. 2(a) in all respects except for the presence of doped silicon tubs $T_S$ and $T_D$ under the source and drain respectively. The relative potential profile of the reference device plotted along the cutline Z-Z' shows the channel potential as expected as shown in Fig. 2(b). Absence of potential wells in source and drain is clearly observed. The transfer characteristics of PWFDSOI MOSFET and the reference device at 10 nm gate length are shown in Fig. 3.

The electric field profile in PWFDSOI MOSFET is altered significantly in comparison to the electric field profile of the reference device as shown in Fig. 4. This can be attributed to the presence of space charge created by the formation of the p-n junctions which makes the presence of the GP more effective in the termination of field lines. This also explains the improvement of DIBL in PWFDSOI MOSFET in comparison to GP as discussed later. Figure 5 shows the output characteristics of 10 nm gate length PWFDSOI MOSFET with BOX thickness of 10 nm with increasing $V_{GS}$ in the absence and presence of back-bias. The phenomenon of velocity saturation is clearly observed in the output characteristics. The performance parameters of 10 nm gate length

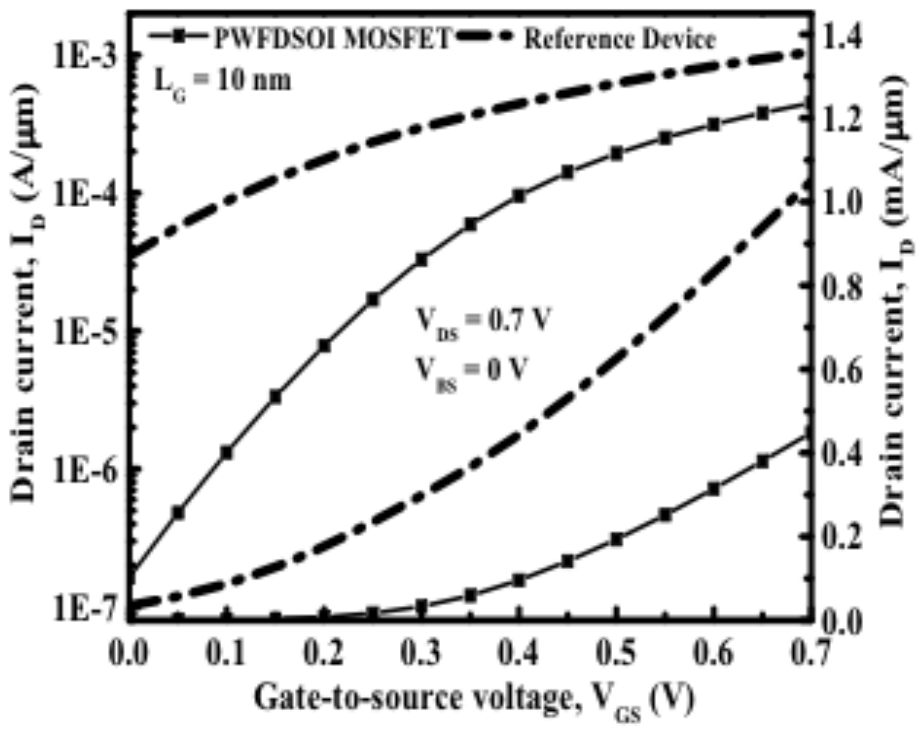

Fig. 3: Transfer characteristics of 10 nm PWFDSOI MOSFET and 10 nm reference device. $V_{DS}$ = 0.7 V and $V_{BS}$ = 0 V.

PWFDSOI MOSFET with a BOX thickness of 10 nm are given in Table 3 for $V_{DS}$ of 0.7 V and $V_{BS}$ of -1.0 V.

## 3. Interpretation of Results

### 3.1. *Drain Induced Barrier Lowering (DIBL)*

The presence of highly doped regions $T_S$ and $T_D$ in PWFDSOI MOSFET significantly reduces the electrostatic coupling between source and drain, and hence, reduces DIBL. Figure 6 shows the determination of threshold voltage using intercept method to calculate DIBL. There is 50% reduction in DIBL in PWFDSOI MOSFET over the reference device as shown in Table 4.

### 3.2. *Potential well depth as function of distance along Y-Y′ in the source*

The variation of depth of potential well in the source as we move from source towards source/$T_S$

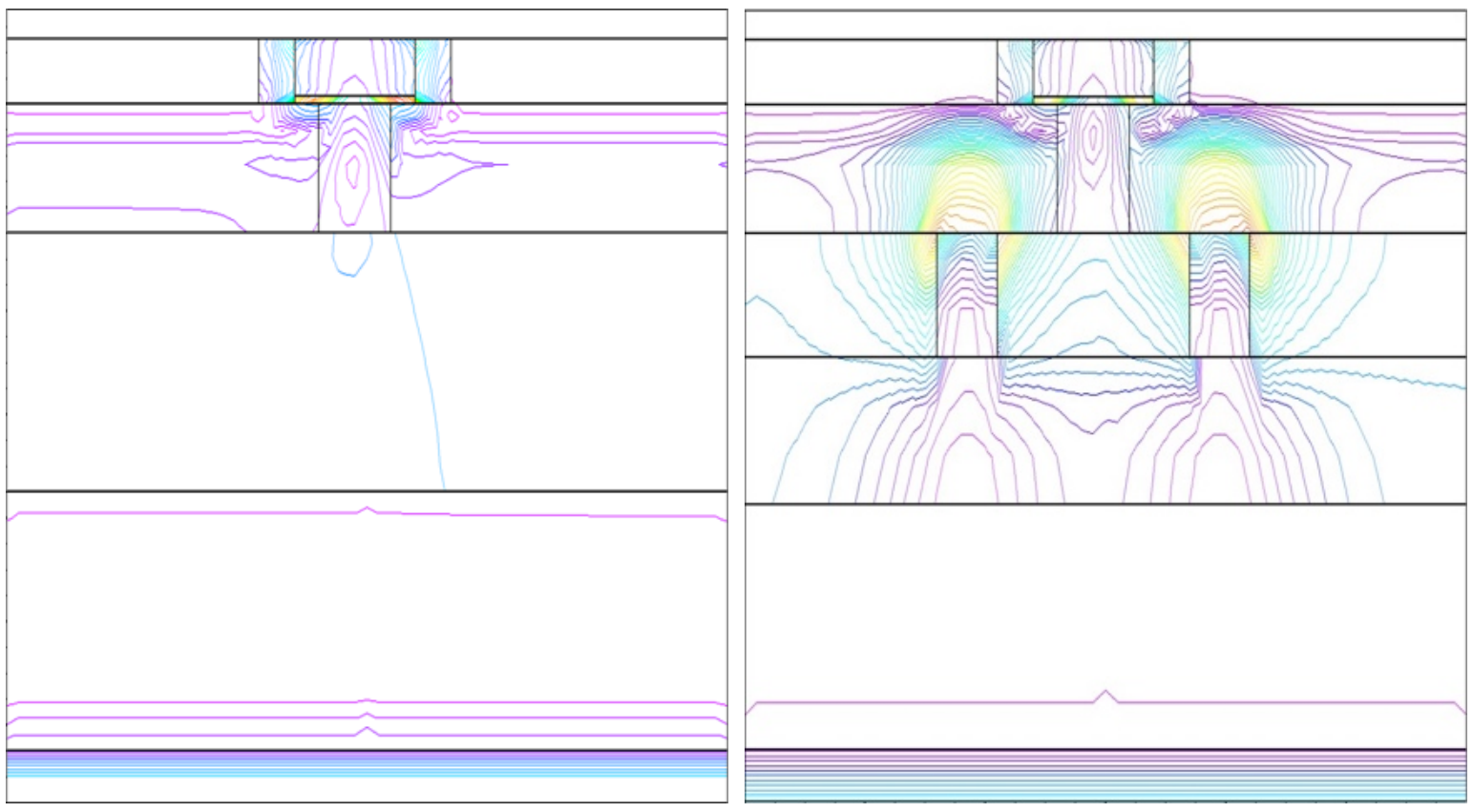

(a) (b)

Fig. 4: Electric field profiles of (a) reference device and, (b) PWFDSOI MOSFET. $V_{DS}$ = 50 mV, $V_{GS}$ = 0 V and $V_{BS}$ = 0 V.

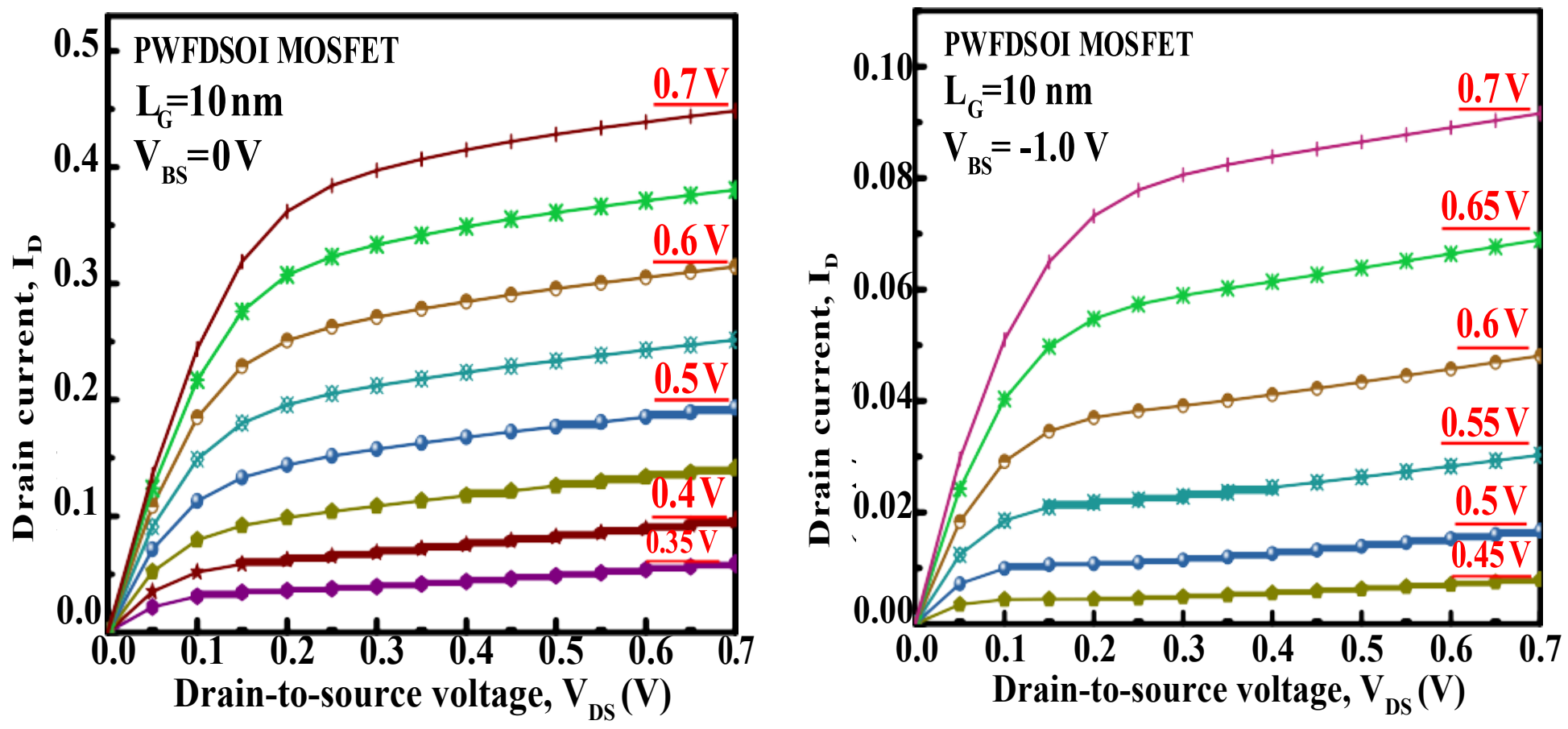

(a) (b)

Fig. 5: $I_D$ vs. $V_{DS}$ characteristics of 10 nm PWFDSOI MOSFET with regions $T_S$ and $T_D$ under the source and drain respectively for (a) $V_{BS}$ = 0 V, and, (b) $V_{BS}$ = -1.0 V. The underlined numbers represent the corresponding value of $V_{GS}$.

Table 3: Performance parameters at 10 nm gate length

| | **PWFDSOI MOSFET** | **Reference Device** | **FinFET (7 nm node)** |
|---|---|---|---|
| SS (mV/decade) | 87 | 144 | 68 |
| $I_{OFF}$ (pA/$\mu$m) | 120 | 3.17 x $10^6$ | 30 |
| $I_{ON}$ ($\mu$A/$\mu$m) | 91.6 | 800 | 285 |
| $I_{ON}/I_{OFF}$ | 7.6 x $10^5$ | 252 | 9.5 x $10^6$ |

Table 4: DIBL values at 10 nm gate length

| | **PWFDSOI MOSFET** | **Reference Device** |
|---|---|---|
| DIBL (mV/V) | 30 | 61 |

interface was studied both in the absence and presence of back-bias. The potential well depth increases as we move along cutline Y-Y' from a point in the source to the source/$T_S$ interface. Figure 7 shows this behaviour. This can be attributed to the increased influence of the positive space charge on the carriers. Thus, electrons located deeper in the source are less likely to contribute to the OFF current. With the application of a back-bias, the potential well depth increases further causing a significant reduction in leakage current as shown in Fig. 8.

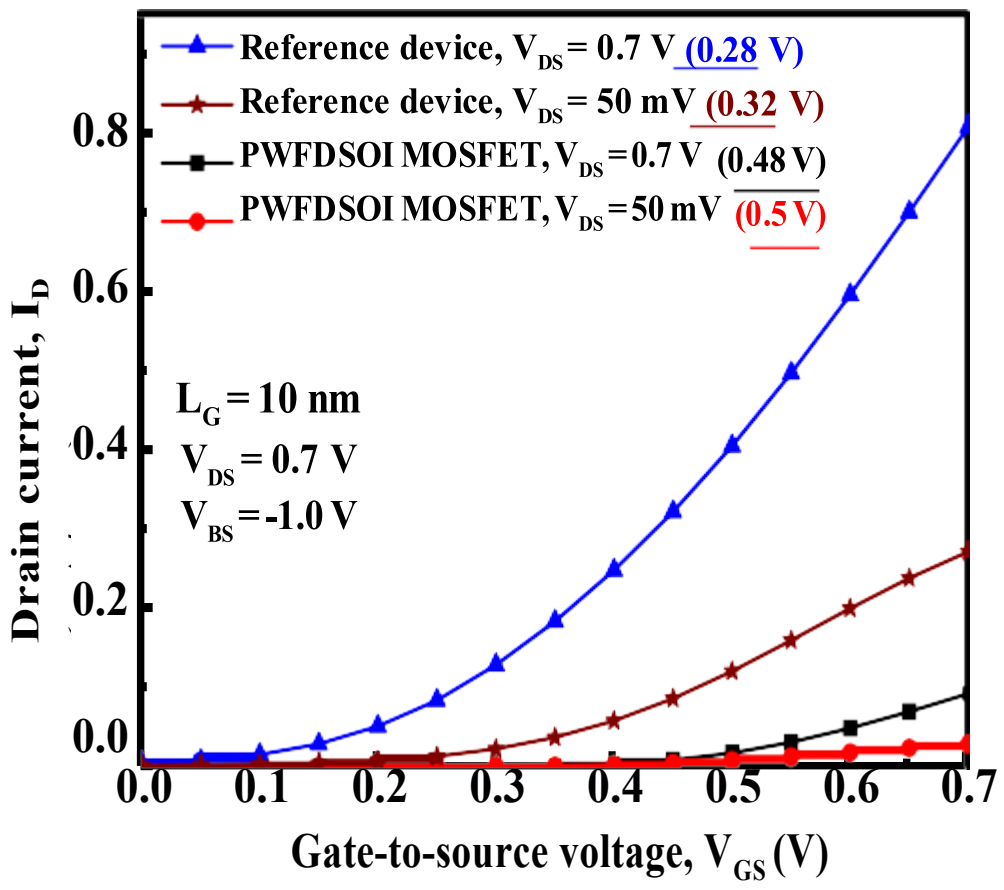

Fig. 6: Transfer characteristics of 10 nm PWFDSOI MOSFET and 10 nm reference device on linear scale to determine threshold voltage using intercept method. The underlined numbers represent the corresponding threshold voltage values. $V_{DS}$ = 0.7 V and $V_{BS}$ = -1.0 V.

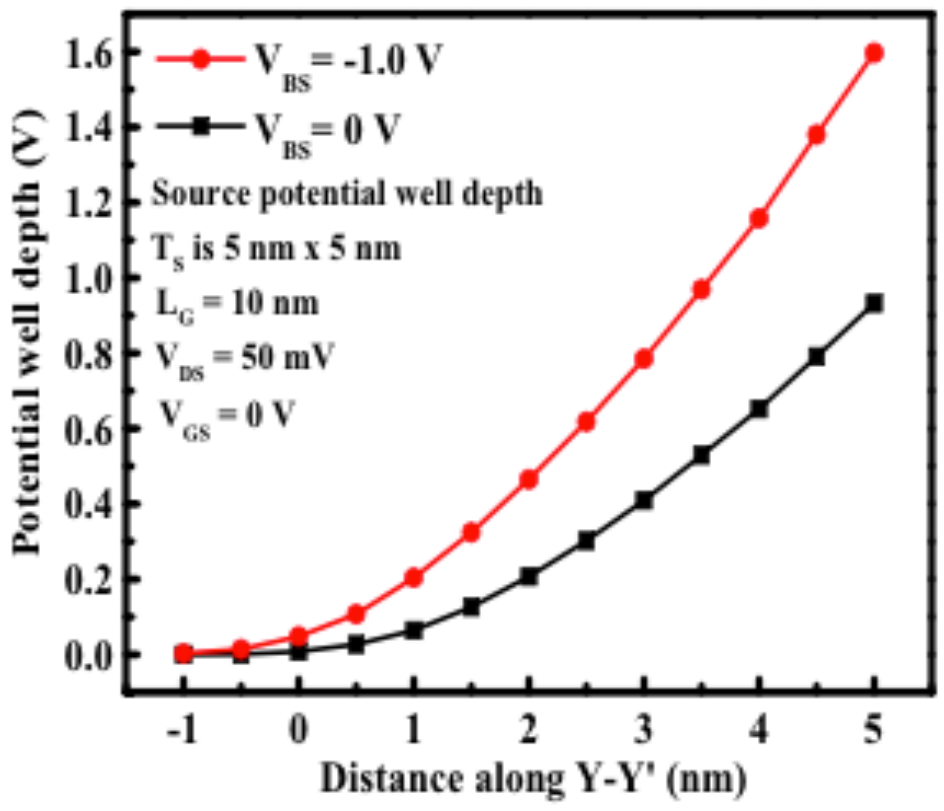

Fig. 7: Variation of potential well depth in the source in 10 nm PWFDSOI MOSFET along cutline Y-Y'. Here Y-Y' =0 is the gate stack/channel interface as shown in Fig. 1(a).

### 3.3. *Potential variation along Y-Y′*

The variation of relative potential was also studied along the cutline Y-Y' across the n-p junction formed by the source and $T_S$ in 10 nm PWFDSOI MOSFET and is shown in Fig. 9 for PWFDSOI MOSFET with BOX thickness of 10 nm. The study was first performed under equilibrium condition ($V_{DS}$=$V_{GS}$=$V_{BS}$=0 V). The difference in the potential across the n-p junction is approximately equal to 1.12 eV which is the band gap energy of silicon as shown in Fig. 9(b). Also, it is clearly observed in Fig. 9 that a significantly larger potential drop occurs in the source region and only about 2 nm depth of $T_S$ from the source/$T_S$ interface is depleted. This suggests significant depth of $T_S$ region is quasi neutral. This implies that process variations in the depth of the $T_S$ and $T_D$ will not make a significant impact on the performance of the device.

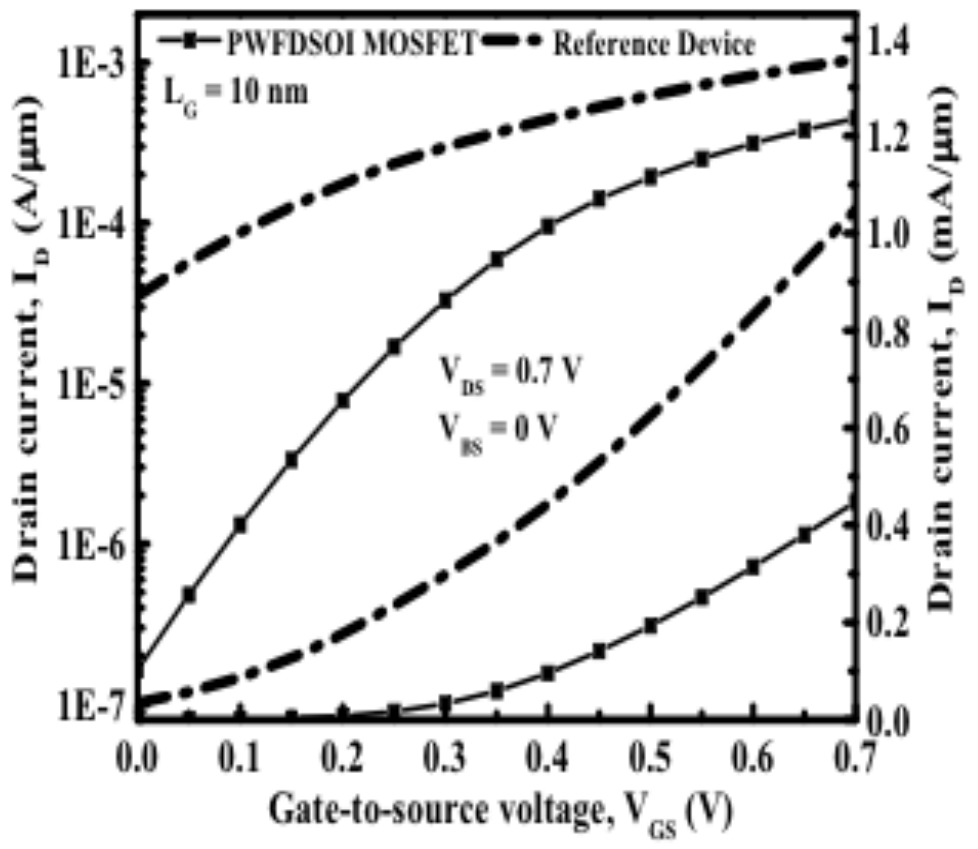

Fig. 8: Transfer characteristics of 10 nm PWFDSOI MOSFET and 10 nm reference device. $V_{DS}$ = 0.7 V and $V_{BS}$ = -1.0 V.

Figure 10 shows the variation of relative potential from source to substrate in 10 nm PWFDSOI MOSFET with 10 nm BOX when a back-bias of -1.0 V is applied to the device. The difference in the potential across the n-p junction under an applied back-bias is approximately equal to Eg/q + $V_{BS}$.

### 3.4. *Effect of Tunneling*

In the proposed device tunneling through the potential wells is of concern primarily when the device is in the OFF state ($V_{GS}$=0 V). To study the effect of tunneling the width of $T_S$ and $T_D$ was extended to the left and right respectively in steps of 1 nm and $I_{OFF}$ was monitored for various values of $T_S$ and $T_D$ width. No appreciable change in $I_{OFF}$ was

145

observed as width of $T_S$ or $T_D$ was varied when BOX thickness was 10 nm under different drain bias conditions as shown in Fig. 11. The OFF current remained fairly constant after a width of 7 nm.

The impact of dopant density of $T_S$ and $T_D$ on $I_{OFF}$ was also studied when BOX thickness was 10 nm. The effect of increased doping of $T_S$ and $T_D$ on $I_{OFF}$ for different widths of $T_S$ and $T_D$ is shown in Fig. 12 at $V_{DS}$ of 50 mV. The OFF current reduces with increased doping density of $T_S$ and $T_D$ regions. However, no effect of increase in width of these regions on $I_{OFF}$ is observed beyond a width of 7 nm. This shows that tunneling is not a significant contributor to the leakage current in PWFDSOI MOSFET. However, for reasons of optimization of $I_{ON}$ /$I_{OFF}$ ratio, a width of 5 nm was chosen for $T_S$ and $T_D$ regions.

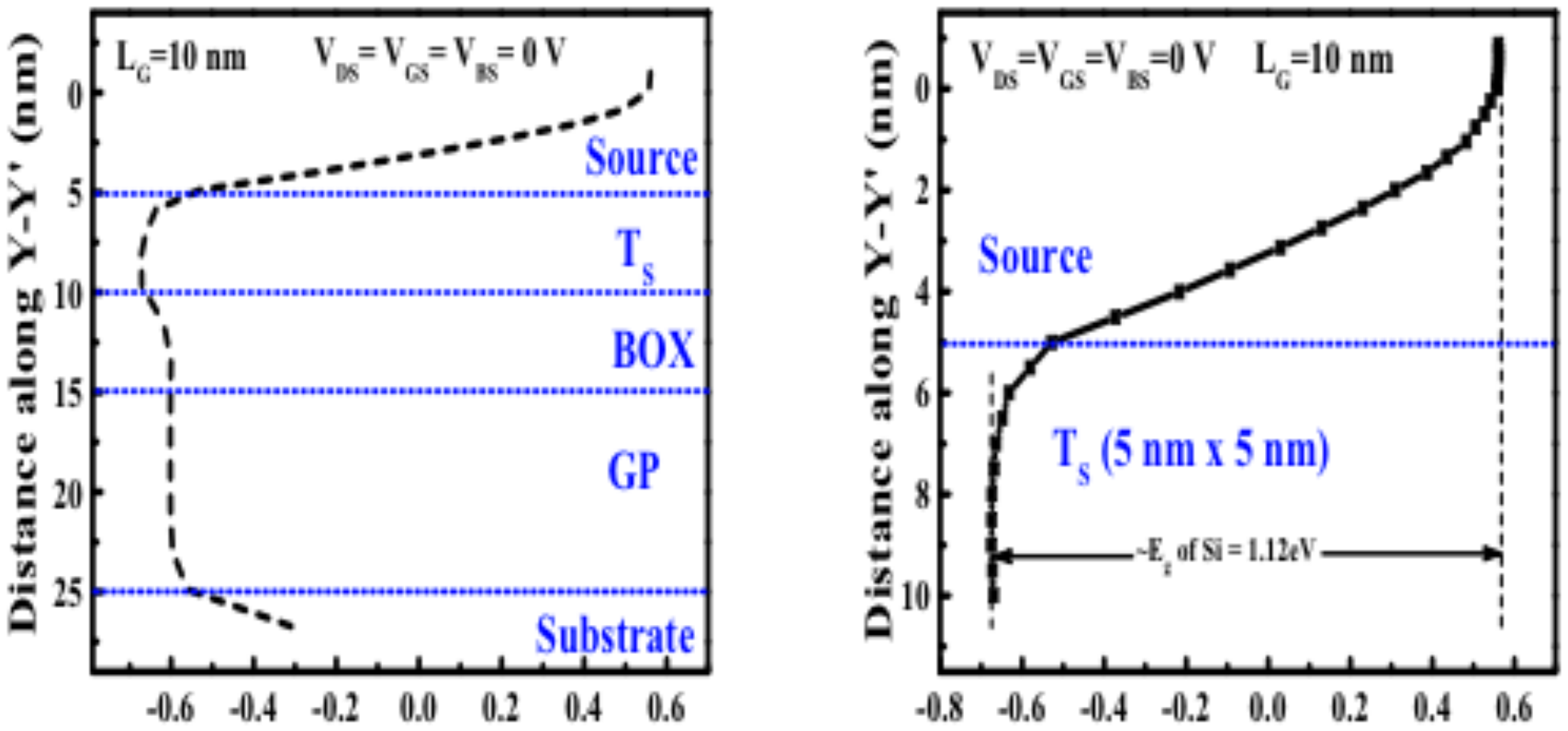

(a) (b)

Fig. 9: Variation of relative potential along cutline Y-Y' (a) from source to substrate, and, (b) from source to $T_S$, in 10 nm PWFDSOI MOSFET with 10 nm thick BOX under equilibrium condition ($V_{DS}$ = $V_{GS}$ = $V_{BS}$ = 0 V). Here, Y-Y' = 0 is the gate stack/channel interface.

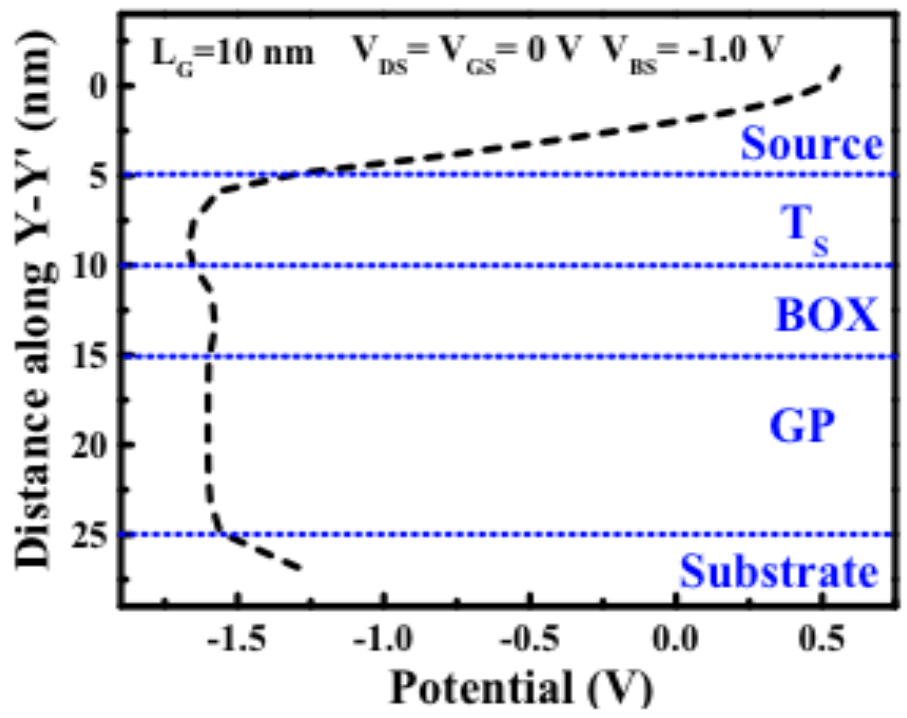

Fig. 10: Variation of relative potential from source to substrate in 10 nm PWFDSOI MOSFET with 10 nm BOX under the condition: $V_{DS}$ = $V_{GS}$ = 0 V, $V_{BS}$ = -1.0 V for doped silicon $T_S$ and $T_D$. Here Y-Y' =0 is the gate stack/channel interface as shown in Fig. 1(a).

### 3.5. *Effect of Dopant Diffusion*

A simulation study was also performed to investigate the impact of dopant diffusion across the source/$T_S$ and drain/$T_D$ interfaces in 10 nm PWFDSOI MOSFET. The diffusion of dopants across the interfaces caused a reduction in depth of the potential wells in the source and drain [1]. Although the potential well depth reduced, the impact on the behavior of the device was not significant as shown in Fig. 13.

## 4. Effect of Increasing BOX Thickness to 15 nm

The schematic of 10 nm gate length PWFDSOI MOSFET with a BOX thickness of 15 nm is identical to Fig. 1(a) in all respects except the BOX thickness. The potential wells in the source and drain regions in the OFF state ($V_{DS}$ = 50 mV, $V_{GS}$ = 0 V, $V_{BS}$ = 0 V) and ON state ($V_{DS}$ = 50 mV, $V_{GS}$ = 0.7 V, $V_{BS}$ = 0 V) of the device along the cutline Z-Z' are same as those for 10 nm thick BOX as dopings of the source/drain regions and tubs is same in both the devices.

### 4.0.1. *Transfer Characteristics*

The $I_D$ vs. $V_{GS}$ characteristics at $V_{DS}$ of 0.7 V and $V_{BS}$ of -1.0 V are shown in Fig. 14. A significant reduction in OFF current is observed due to the presence of $T_S$ and $T_D$. The device performance parameters for the case of 15 nm thick BOX are given in Table 5. PWFDSOI MOSFET with 15 nm thick BOX is significantly better than the reference device having same BOX thickness.

### 4.0.2. *Electric field Profile and Potential Contour*

The electric field profile and potential contour of PWFDSOI MOSFET with BOX thickness of 15 nm are shown in Fig. 15. The electric field profile in PWFDSOI MOSFET is significantly differ- ent from that of the reference device under identi- cal bias conditions and causes 50% reduction in DIBL in PWFDSOI MOSFET as compared to the reference GP FDSOI MOSFET.

### 4.0.3. *Relative Potential along Y-Y' upto Substrate*

The relative potential profile along Y-Y' upto the substrate for PWFDSOI MOSFET with 15 nm thick BOX in the presence of a back-bias of -1.0 V is shown in Fig. 16. The difference in the potential

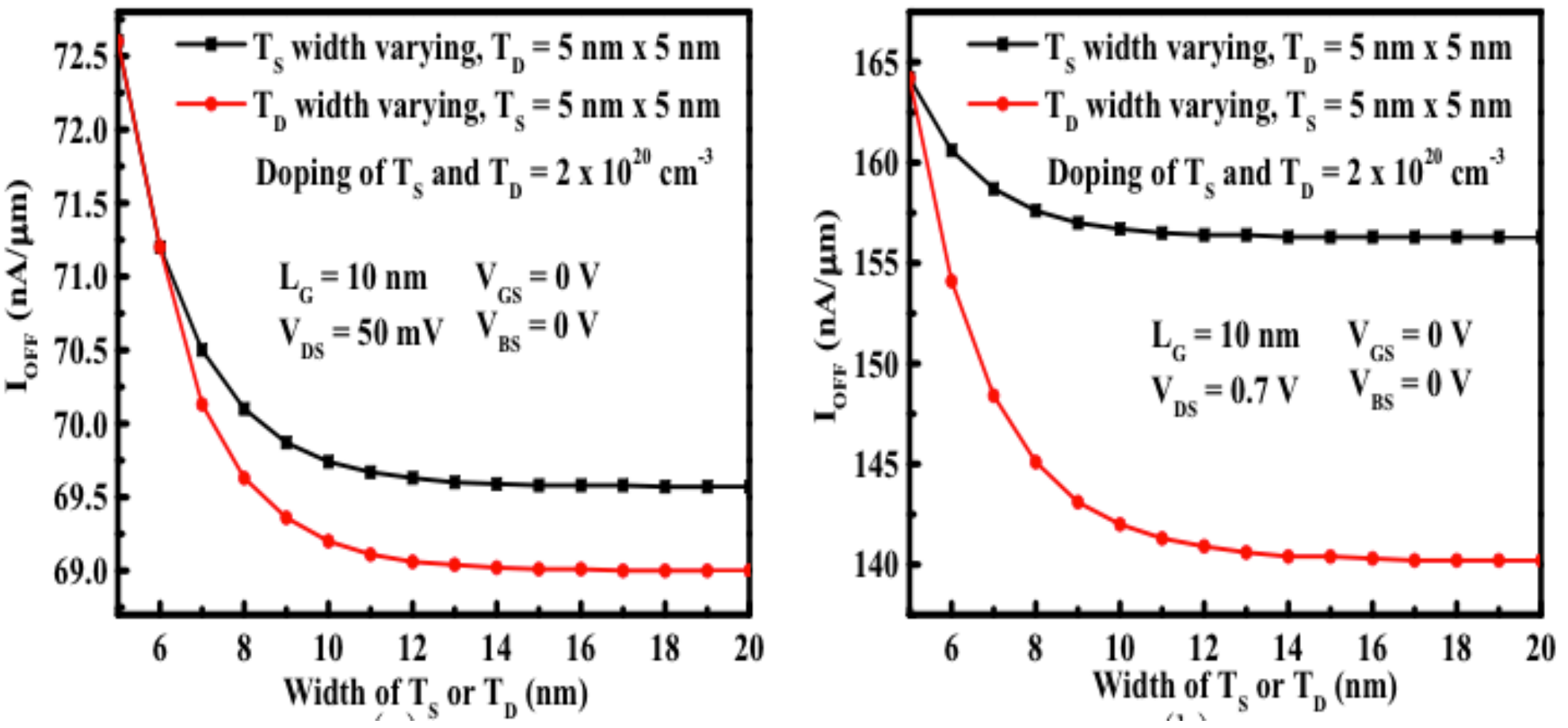

(a) (b)

Fig. 11: Variation of $I_{OFF}$ with the width of $T_S$ and $T_D$ in the proposed 10 nm PWFDSOI MOSFET (a) at $V_{DS}$=50 mV, and (b) at $V_{DS}$ =0.7 V. Doping of $T_S$ and $T_D$ = 2 x $10^{20}$ $cm^{-3}$, $V_{GS}$ = 0 V and $V_{BS}$ = 0 V.

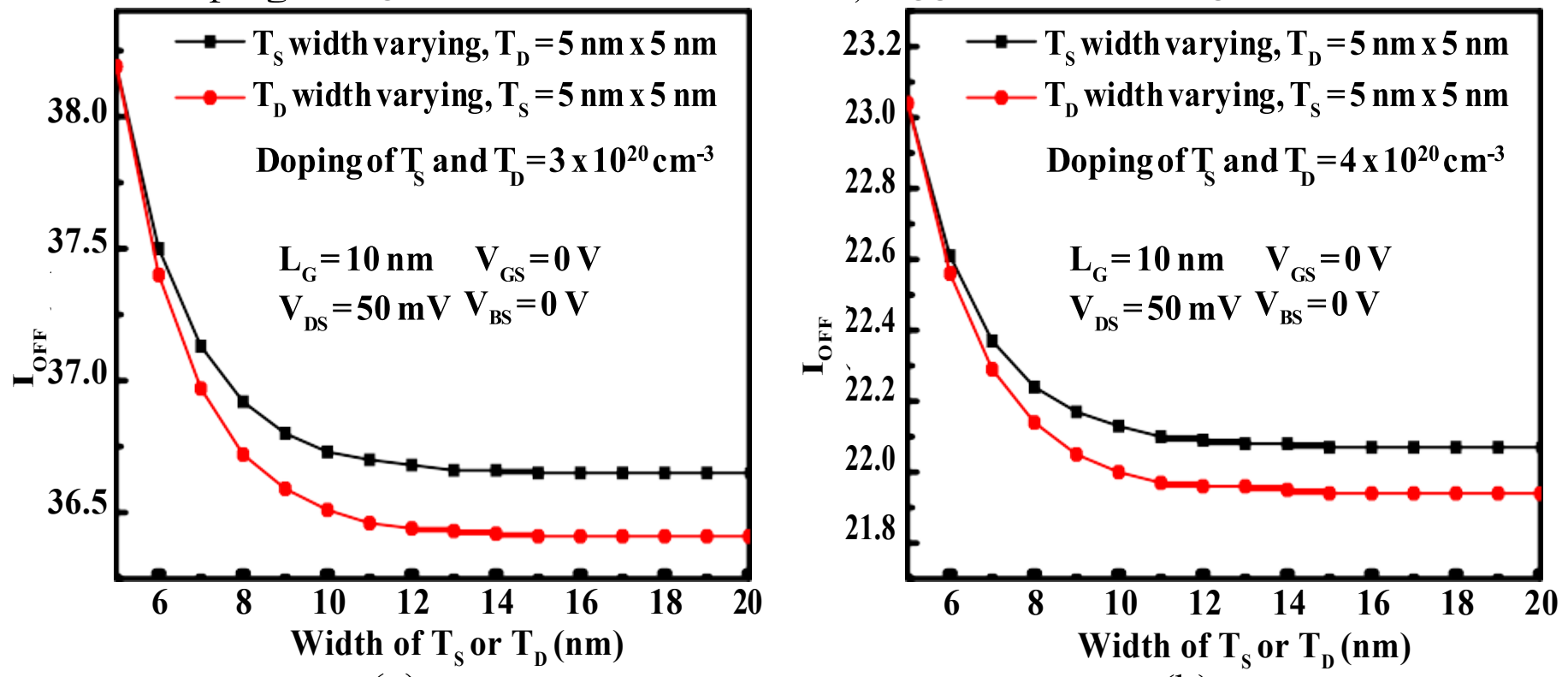

(a) (b)

Fig. 12: Variation of $I_{OFF}$ with the width of $T_S$ and $T_D$ in the proposed 10 nm PWFDSOI MOSFET at $V_{DS}$ =50 mV when doping of $T_S$ and $T_D$ is (a) 3 x $10^{20}$ $cm^{-3}$, and, (b) 4 x $10^{20}$ $cm^{-3}$. $V_{GS}$ = 0 V and $V_{BS}$ = 0 V.

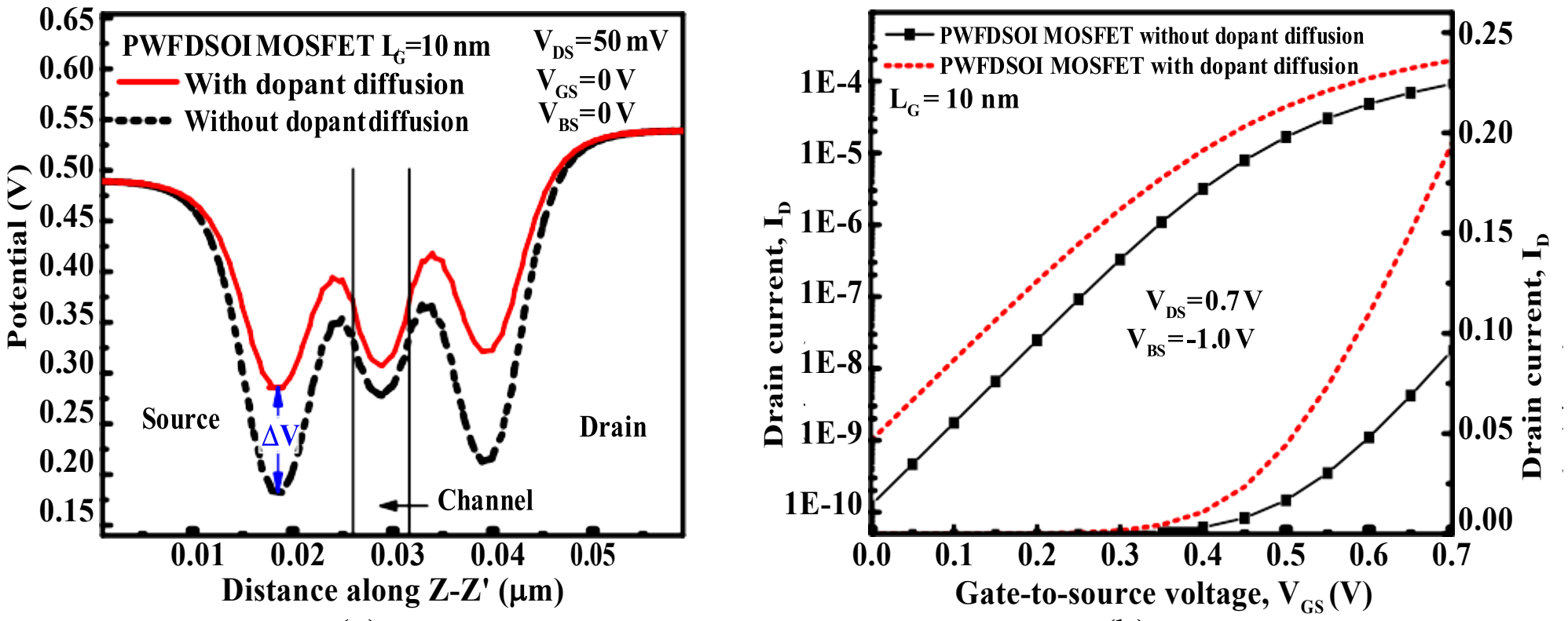

(a) (b)

Fig. 13: (a) Reduction in potential well depth with dopant diffusion across source/$T_S$ and drain/$T_D$ interfaces. $V_{DS}$ = 50 mV, $V_{GS}$ = $V_{BS}$ = 0 V. (b) Transfer characteristics of PWFDSOI MOSFET when dopant diffusion occurs across source/$T_S$ and drain/$T_D$ interfaces.

Table 5: Performance parameters at 10 nm gate length with BOX thickness of 15 nm

| | PWFDSOI MOSFET | Reference Device |
|---|---|---|
| SS (mV/decade) | 87 | 190 |
| $I_{OFF}$ (pA/µm) | 144 | 18.32 x $10^6$ |
| $I_{ON}$ (µA/µm) | 91.5 | 980 |
| $I_{ON}$/$I_{OFF}$ | 6.3 x $10^5$ | 53 |

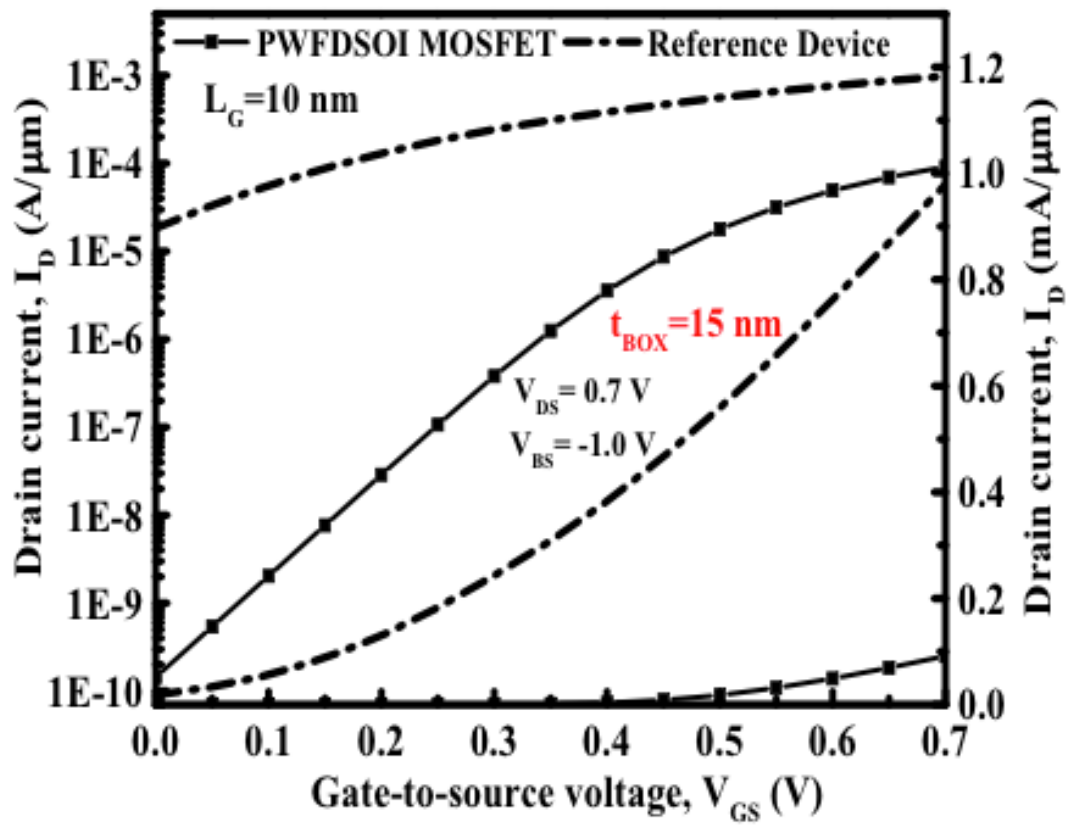

Fig. 14: $I_D$ vs. $V_{GS}$ characteristics of 10 nm gate length PWFDSOI MOSFET with BOX thickness of 15 nm. $V_{DS}$ = 0.7 V, $V_{BS}$ = -1.0 V.

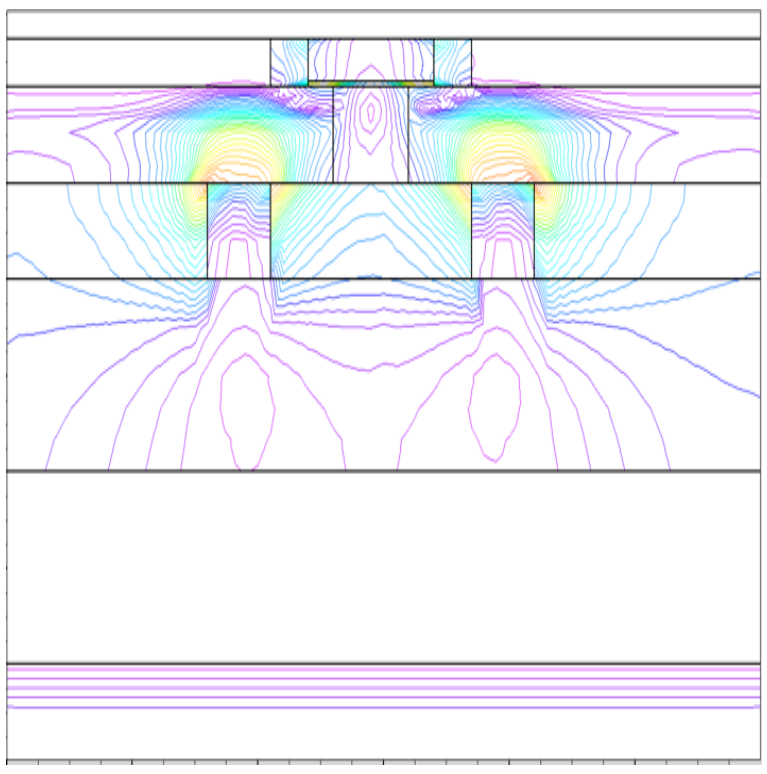

(a)

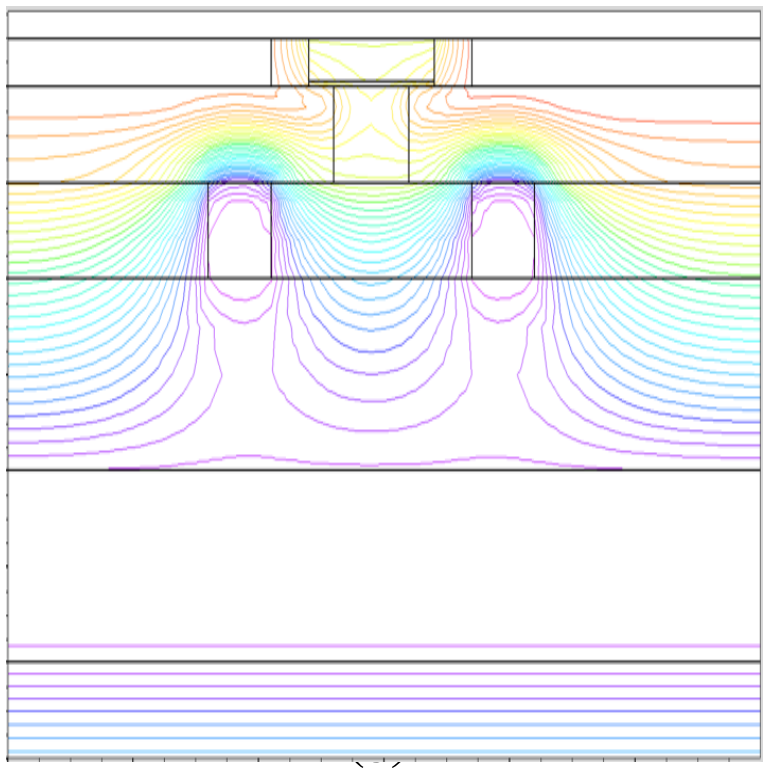

Fig. 15: (a) Electric field profile and, (b) Potential contour of PWFDSOI MOSFET with BOX thickness of 15 nm. $V_{DS}$ = 50 mV, $V_{GS}$ = 0 V and $V_{BS}$ = 0 V.

across the n-p junction formed by source and $T_S$ is approximately equal to $E_g/q + V_{BS}$.

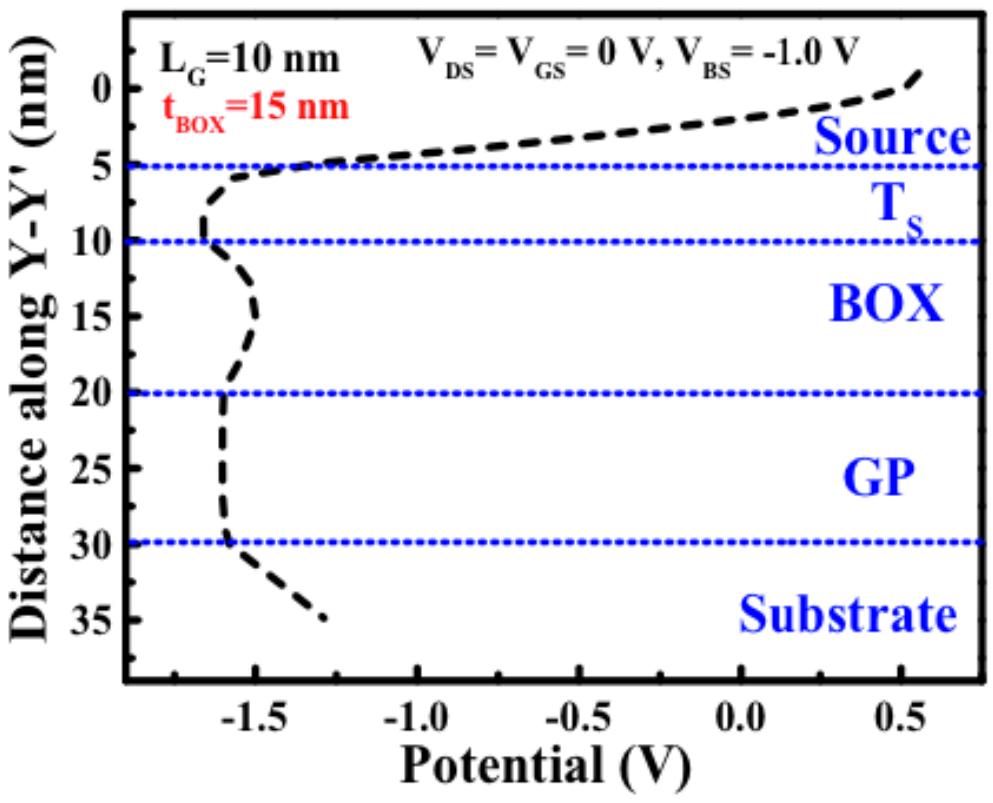

Fig. 16: Variation of relative potential from source to substrate in 10 nm PWFDSOI MOSFET with 15 nm thick BOX. Here, Y-Y' = 0 is the gate stack/channel interface.

## 5. Conclusion

A simulation study to get an insight into the physics of PWFDSOI MOSFET was presented in this paper. This study was performed on devices with unstrained silicon channel. The scalability to 10 nm gate length was discussed in detail. DIBL in PWFDSOI MOSFET was found to be significantly lower as compared to the GP FDSOI MOSFET due to the altered electric field. The effect of dopant diffusion and tunneling through the wells was also discussed. It was also found that the depth of the doped tubs is largely insensitive to variations during processing. Effect of increasing BOX thickness to 15 nm was also studied.